\documentclass[10pt,a4paper]{book}
\usepackage{frontier04}
\begin{document}
\newcounter{ctr}
\setcounter{ctr}{\thepage}
\talktitle{Structure formation, CMB and LSS in a mirror dark matter scenario}
\talkauthors{Paolo Ciarcelluti \structure{a,b}}

\begin{center}
\authorstucture[a]{Dipartimento di Fisica, 
                              Universit\`a di L'Aquila,  \newline
                              67010 Coppito AQ, Italy}

\authorstucture[b]{INFN, Laboratori Nazionali del Gran Sasso, 
                              67010 Assergi AQ, Italy}
\end{center}
\shorttitle{Structure formation, CMB and LSS in a mirror ...} 
\firstauthor{Paolo Ciarcelluti}

\begin{abstract}
In the mirror world hypothesis the mirror baryonic component emerges
as a possible dark matter candidate. 
Here we study the behaviour of the mirror dark matter and the differences 
from the more familiar CDM candidate for structure formation, 
cosmic microwave background and large scale structure.
We show mirror models for CMB and LSS power spectra and compare them 
with observations, obtaining bounds on the mirror parameter space.
\end{abstract}

\section{Introduction}

The idea that there may exist a hidden mirror sector 
of particles and interactions with exactly the same properties  
as our visible world 
was suggested long time ago by Lee and Yang \cite{mirror}, and 
the model with exact parity symmetry interchanging corresponding fields 
of two sectors was proposed many years later by Foot at al. \cite{mirror}.
The two sectors communicate with each other only via gravity\footnote{
There could be other interactions, as for example the kinetic mixing 
between O and M photons \cite{mixing}, but they are negligible for the 
present study.}. 
A discrete symmetry $G\leftrightarrow G'$ interchanging
corresponding fields of $G$ and $G'$, so called mirror parity,  
guarantees that two particle sectors are described by 
identical Lagrangians, with all coupling constants 
(gauge, Yukawa, Higgs) having the same pattern. 
As a consequence the two sectors should have the same 
microphysics\footnote{Here we do not consider 
the possibility that the mirror parity is spontaneously broken, leading to 
somewhat different particle physics in the mirror sector \cite{broken}.}. 
After its first applications to non-baryonic dark matter 
\cite{blinkhlo}, the mirror matter hypothesis has been invoked in 
many physical and astrophysical questions: 
large scale structure of the Universe \cite{ignavol-lss,paolo}, 
galactic halo \cite{mir_halo},
MACHOs \cite{Macho}, 
gamma ray bursts \cite{mir_GRB}, 
orthopositronium lifetime \cite{ortho,bader}, 
neutrino physics \cite{neutrino}, 
interpretation of dark matter detection experiments \cite{mir_dama}, 
meteoritic event anomalies \cite{mir_meteor,detect_mir_frag}, 
close-in extrasolar planets \cite{mir_planet}, 
Pioneer spacecraft anomalies \cite{pioneer}.

If the mirror (M) sector exists, then the Universe
along with the ordinary (O) particles 
should contain their mirror partners, but their densities are not the same 
in both sectors. 
In fact, Berezhiani et al. \cite{bcv} showed that the BBN bound on the 
effective number of extra light neutrinos implies that the M sector has 
a temperature lower than the O one, that can be naturally achieved in 
certain inflationary models \cite{inflation}. 
Then, two sectors have different initial conditions, they do not come into 
thermal equilibrium at later epoch and they evolve independently,
maintaining  approximately constant the ratio among their temperatures. 

All the differences with respect to the ordinary world can be described in 
terms of only two free parameters in the model, $ x \equiv T' / T $ and 
$ \beta \equiv \Omega'_{\rm b} / \Omega_{\rm b} $, where $T$ ($T'$) and 
$\Omega_{b}$ ($\Omega'_{b}$) are respectively the ordinary (mirror) photon 
temperature and the ordinary (mirror) baryon density. 
The bounds on the mirror parameters are $ x < 0.64 $ \cite{bcv} and 
$ \beta > 1 $, the first one coming from the BBN limit and the second one 
from the hypothesis that a relevant fraction of dark matter is made of mirror 
baryons.

In fact, if $\Omega'_b \geq \Omega_b$, mirror baryons emerge as a 
possible dark matter candidate (MBDM) \cite{baryo-lepto};
the peculiar properties of mirror dark matter were discussed
in ref. \cite{paolo}.

\section{Relevant length scales}

The important moments for the structure formation are related to the 
matter-radiation equality (MRE) and to the matter-radiation decoupling 
(MRD) epochs. 
The MRE occurs at the redshift
\begin{equation} \label{z-eq} 
1+z_{\rm eq}= {{\Omega_m} \over {\Omega_r}} \approx 
 2.4\cdot 10^4 {{\Omega_{m}h^2} \over {1+x^4}} \;.
\end{equation}
Therefore, for $x\ll 1$ it is not altered by the additional relativistic 
component of the M sector.
The mirror MRD temperature $T'_{\rm dec}$ 
can be calculated following the same lines as in 
the O one \cite{bcv}, and hence 
\begin{equation} \label{z'_dec}
1+z'_{\rm dec} \simeq x^{-1} (1+z_{\rm dec}) 
\simeq 1100 \; x^{-1} \; ,
\end{equation}
so that the MRD in the M sector occurs earlier than in the O one. 
Moreover, for values 
$ x < x_{\rm eq} \simeq 0.046 \, \left( \Omega_{m} h^2 \right)^{-1}$, 
the mirror photons would decouple yet during the radiation dominated 
period. 
This critical value plays an important role in our further considerations, 
where we distinguish between two cases: $x > x_{\rm eq}$ and 
$x < x_{\rm eq}$. 

The relevant scale for gravitational instabilities is the mirror Jeans mass, 
defined 
as the minimum scale at which, in the matter dominated epoch, sub-horizon 
sized perturbations start to grow. 
In the case $x > x_{\rm eq}$ (where the mirror decoupling happens after the 
matter-radiation equality) its maximum value is reached just before 
the mirror decoupling, and is expressed in terms of the O one as
\begin{equation}
M_{\rm J,max}' \approx 
  \beta^{-1/2} \left( { x^4 \over {1 + x^4} } \right)^{\rm 3/2} 
  \cdot M_{\rm J,max} \;,
\end{equation}
which, for $\beta \geq 1$ and $x < 1$, means that the Jeans mass for the M 
baryons is lower than for the O ones, 
with implications for the structure formation process. 
If, e.g., $ x = 0.6 $ and $ \beta = 2 $, then 
$ M_{\rm J}' \sim 0.03 \; M_{\rm J} $. 
We can also express the same quantity in terms of $ \Omega_b $, $ x $ 
and $ \beta $, in the case that all the dark matter is in the form of mirror 
baryons, as
\begin{equation} \label{mj_mir_1}
M_{\rm J}'(a_{\rm dec}') \approx 
  3.2 \cdot  10^{14} M_\odot \;
  \beta^{-1/2} ( 1 + \beta )^{-3/2} \left( x^4 \over {1+x^4} \right)^{3/2} 
  ( \Omega_{\rm b} h^2 )^{-2} \;.
\end{equation}
For the case $ x < x_{\rm eq} $, the mirror decoupling happens before the 
matter-radiation equality.
In this case we obtain for the highest value of the Jeans mass just 
before decoupling the expression
\begin{equation}
 \label{mj_mir_2}
M_{\rm J}'(a_{\rm dec}') \approx 
  3.2 \cdot  10^{14} M_\odot \; 
  \beta^{-1/2} ( 1 + \beta )^{-3/2} 
  \left( x \over x_{\rm eq} \right)^{3/2} \left( x^4 \over {1+x^4} \right)^{3/2} 
  ( \Omega_{\rm b} h^2 )^{-2} \;.
\end{equation}
In case $ x = x_{\rm eq} $, the expressions 
(\ref{mj_mir_1}) and (\ref{mj_mir_2}), respectively valid for 
$ x \ge x_{\rm eq} $ and $ x \le x_{\rm eq} $, are coincident, as we expect.
If we consider the differences between the highest mirror Jeans mass for 
the particular values $ x = x_{\rm eq}/2 $, $ x = x_{\rm eq} $ 
and $ x = 2 x_{\rm eq} $, we obtain the following relations:
\begin{equation}
M_{\rm J,max}'(x_{\rm eq}/2) \approx 0.005 \: M_{\rm J,max}'(x_{\rm eq}) \;,
\end{equation}
\begin{equation}
M_{\rm J,max}'(2x_{\rm eq}) \approx 64 \: M_{\rm J,max}'(x_{\rm eq}) \;.
\end{equation}
Density perturbations in MBDM on scales $ M \ge M'_{\rm J,max}$ which 
enter the horizon at $z\sim z_{\rm eq}$ undergo uninterrupted linear 
growth. 
Perturbations on scales $ M \le M'_{\rm J,max}$ start instead to oscillate 
after they enter the horizon, thus delaying their growth till the 
epoch of M photon decoupling.

As occurs for perturbations in the O baryonic sector, 
also the M baryon density fluctuations should undergo the 
strong collisional Silk damping around the time of M recombination, 
so that 
the smallest perturbations that survive the dissipation will have the mass 
\begin{equation} \label{ms_m}
M'_S \sim [f(x) / 2]^3 (\beta \, \Omega_b h^2)^{-5/4} 10^{12}~ M_\odot \;,
\end{equation}
where $f(x)=x^{5/4}$ for $x > x_{\rm eq}$, 
and $f(x) = (x/x_{\rm eq})^{3/2} x_{\rm eq}^{5/4}$ 
for $x < x_{\rm eq}$. 
For $x\sim x_{\rm eq}$ we obtain 
$ M'_S \sim 10^{10} ~M_\odot $, a typical galaxy mass.

\section{CMB and LSS spectra}

In order to obtain quantitative predictions we computed numerically the 
evolution of scalar adiabatic perturbations in a flat Universe in which is 
present a significant fraction of mirror dark matter at the expenses of 
diminishing the CDM contribution and maintaining constant $\Omega_m$. 
\begin{figure}[h]
  \begin{center}
    \epsfig{figure=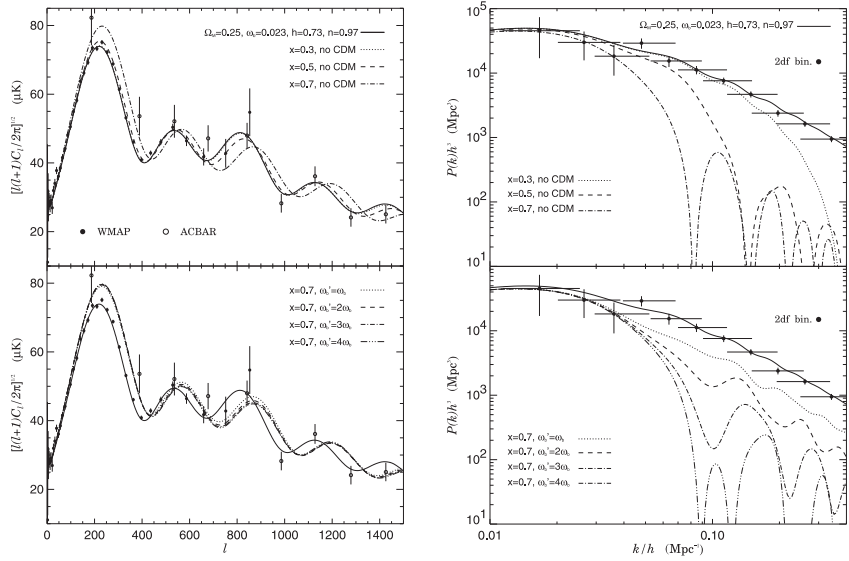,width=11.5cm}  
  \end{center}
\caption{\small CMB ({\sl left}) and LSS ({\sl right}) power spectra for 
different values of $x$ and $\omega_{\rm b}'$, as compared with a reference 
standard model (solid line) and with observations (WMAP \cite{wmap-par}, 
ACBAR \cite{acbar-data} and 2dF binned data \cite{2df-teg}).
Models where dark matter is entirely due to MBDM (no CDM) are plotted 
in {\sl top panels} for $x = 0.3, 0.5, 0.7$, while models with mixed 
CDM+MBDM ($\beta=1,2,3,4$ ; $x=0.7$) in {\sl bottom panels.} }
\label{cmblssfig3}
\end{figure}
We have chosen a  ``reference cosmological model'' 
with the following set of parameters \cite{wmap-par}: 
$ \omega_{b} = 0.023, ~ 
\Omega_{ m} = 0.25, ~ 
\Omega_{\Lambda} = 0.75 , ~ 
n_{\rm s} = 0.97, ~ h = 0.73 $. 

The dependence of the CMB and LSS power spectra on the parameters 
$x$ and $\beta$ is shown in fig. \ref{cmblssfig3}. 
The predicted CMB spectrum is quite strongly dependent on the value of 
$x$, and it becomes practically indistinguishable from the CDM case for 
$x < x_{\rm eq} \approx 0.3$.
However, the effects on the CMB spectrum rather weakly depend on the
fraction of mirror baryons. 
As a result of the oscillations in MBDM perturbation evolution, one observes 
oscillations in the  LSS power spectrum; their position 
clearly depends on $x$, while their depth depends on the mirror baryonic 
density.
Superimposed to oscillations one can see the cut-off in the power 
spectrum due to the aforementioned Silk damping.

In the same figure 
our predictions can be compared with the observational data in order to 
obtain some general bound on the mirror parameters space.

\begin{itemize}

\item The present LSS data are not compatible with a scenario where all the 
dark matter is made of mirror baryons, unless we consider enough small 
values of $ x $: 
$ x \le 0.3 \approx x_{\rm eq} $.

\item High values of $ x $, $ x > 0.5 $, can be excluded
even for a relatively small amount of mirror baryons. 
In fact, we observe 
relevant effects on LSS and CMB power spectra down to values of 
M baryon density of the order $ \Omega'_b \sim \Omega_b $. 

\item Intermediate values of $ x $, $ 0.3 < x < 0.5 $, can be 
allowed if the MBDM is a subdominant component of dark matter, 
$ \Omega'_b \le \Omega_b \le \Omega_{CDM} $. 

\item For small values of $ x $, $ x < 0.3 $, 
the MBDM and the CDM scenarios are indistinguishable as concerns 
the CMB and the linear LSS power spectra.
In this case, in fact, the mirror Jeans and Silk lengths, 
which mark region of the spectrum where the effects of 
mirror baryons are visible, decrease to very low values, which undergo 
non linear growth from relatively large redshift. 

\end{itemize}

Thus, with the current experimental accuracy, we can exclude only 
models with high $ x $ and high $ \Omega_b' $; 
however, there can be many possibilities to disentangle the cosmological 
scenario of two parallel worlds with the future high precision data 
concerning the large scale structure, CMB anisotropy, structure of the 
galaxy halos, gravitational microlensing, oscillation of 
observable particles into their mirror partners, etc.  

\section*{Acknowledgements}

\noindent 
I am grateful to my invaluable collaborators Zurab Berezhiani, Denis 
Comelli and Francesco Villante. 
I would like to thank also Silvio Bonometto, Stefano Borgani, Alfonso 
Cavaliere and Nicola Vittorio for interesting discussions.

\end{document}